\documentclass[%
 reprint,
 amsmath,amssymb,
 aps,
]{revtex4-2}

\usepackage{graphicx}% Include figure files
\usepackage{dcolumn}% Align table columns on decimal point
\usepackage{bm}% bold math
\begin{document}

\preprint{APS/123-QED}

\title{Charge order in the Pr substituted YBa$_2$Cu$_3$O$_7$ from high-field Hall effect measurements}% Force line breaks with \\

\author{C. M. Duffy$^{1,\dagger}$, M. Altangerel$^{1,2,3}$, S. Badoux$^{1}$, D. Vignolles$^{1,3}$, T. Oustric$^{2}$, C.~M.~Moir$^{4}$, Keke Feng$^{4}$, A. Frano$^{4}$, M. B. Maple$^{4}$, L. Taillefer$^{2,3,5}$ and C. Proust$^{1,3,\ddagger}$}

\affiliation{
$^1$Univ. Toulouse, INSA-T, Univ. Grenoble Alpes, CNRS, LNCMI-EMFL, UPR3228, Toulouse, France.\\ 
$^2$Institut quantique, Département de physique \& RQMP, Université de Sherbrooke, Sherbrooke, Québec, Canada\\
$^3$Université de Sherbrooke–CNRS and IRL Frontières Quantiques, Sherbrooke, Québec, Canada.\\
$^4$Department of Physics, Center for Advanced Nanoscience, University of California, San Diego, CA 92093, USA.\\
$^5$Canadian Institute for Advanced Research, Toronto, Ontario M5G 1M1, Canada\\
}

\date{\today} 

\begin{abstract}
The mechanism of doping in the composite Pr$_x$Y$_{1-x}$Ba$_2$Cu$_3$O$_{7-\delta}$ (Pr-YBCO) system is distinct from that of pure YBCO, offering a means to explore the requirements for the numerous electronic orders appearing in the phase diagram. One such example is the ubiquitous 2D charge order and concomitant Fermi surface reconstruction in underdoped YBCO. Here, using magnetotransport and Hall effect measurements, we find signatures of a Fermi surface reconstruction similar to that in pure YBCO indicating the presence of 2D charge order in Pr-YBCO. Additionally, we find that the phase diagrams of Pr-YBCO and YBCO are decidedly symmetric despite the additional disorder in the former and the distinction between hole depletion through Pr substitution and through O reduction. This indicates that while the mechanism of doping differs, the amount of charge carriers in the planes is the most important factor governing the electronic orders in these systems.
\end{abstract}

\maketitle

\section{\label{Intro} Introduction}

\begin{figure}[t]
        \includegraphics[width = 0.45\textwidth]{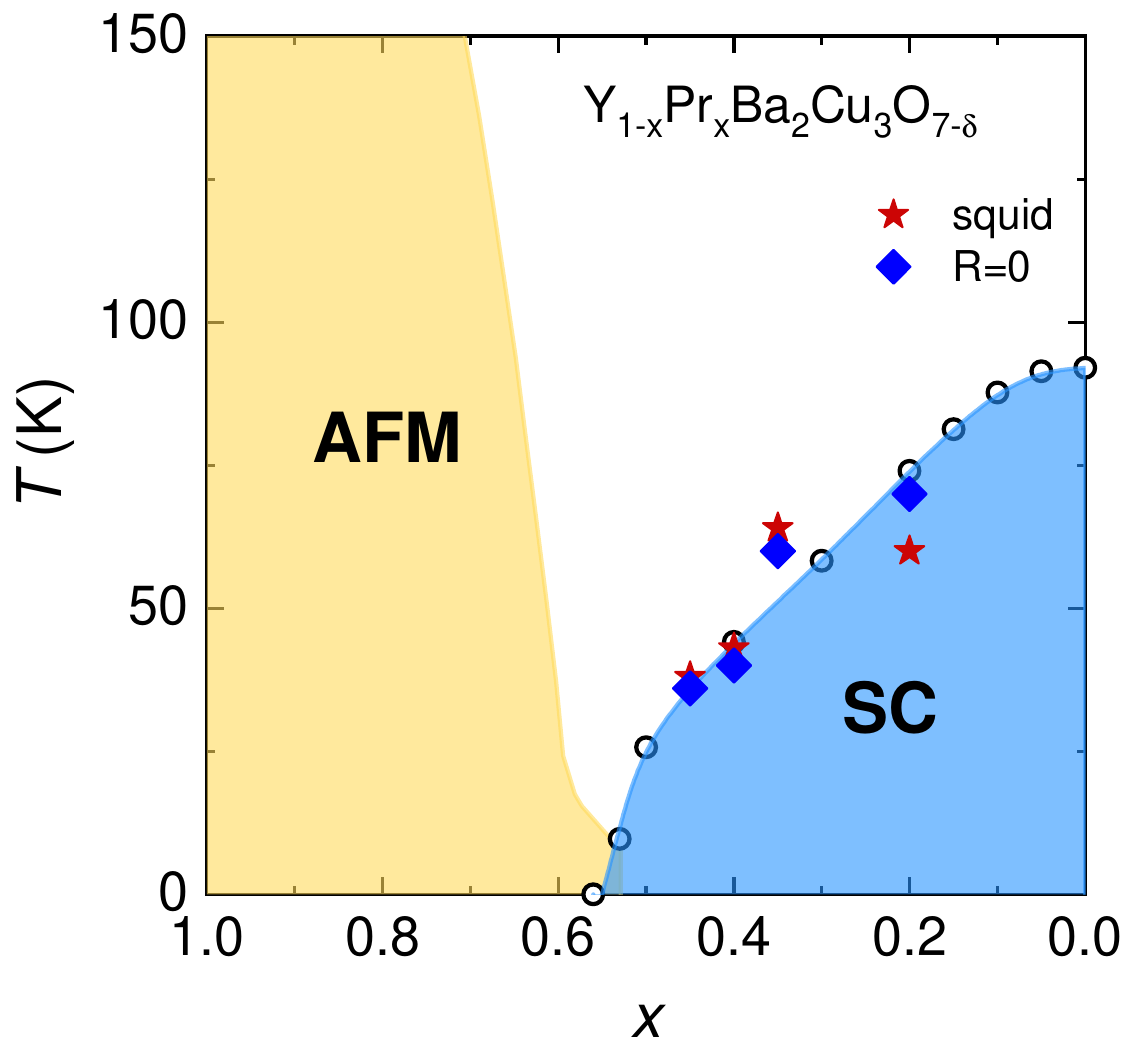}
        \caption{\label{fig:phasediagram} The phase diagram of Pr-YBCO. The yellow region marks the antiferromagnetic (AFM) region bounded by $T_N$. Muon spin-relaxation measurements reveal two types of magnetic order: the suppression of the Mott state, and magnetic order from the Pr spin. The superconducting dome is marked in blue. The star (diamond) symbols correspond to $T_c$ measured by squid (resistivity) in the four Pr-YBCO samples presented in this study. }
    \end{figure}

    The discovery of YBa$_2$Cu$_3$O$_{7-\delta}$ (YBCO) marked a paradigm shift in the field of high-$T_c$ superconductivity. Its high maximum $T_c$ at optimal doping (90 K), cleanliness, and ease of synthesis have made it a key system for studies on high-$T_c$ cuprates. The complete chemical substitution of Y for any rare-earth element (such as Nd, Er, and Gd) has no impact on $T_c$ with a few exceptions: Ce and Tb which do not readily form the orthorhombic crystal structure of YBCO, Pm which is radioactively unstable, and Pr which forms a stable, isostructural system with the caveat of being an insulator \cite{Akhavan_PhysB_2002}. The composite system Pr$_x$Y$_{1-x}$Ba$_2$Cu$_3$O$_{7-\delta}$ (Pr-YBCO), whose phase diagram is presented in Fig. \ref{fig:phasediagram}, demarcates the superconducting and insulating states, offering a unique insight into the role of doping, disorder, and oxygen in the YBCO system. Superconductivity is continually tuned through Pr cation substitution while retaining the oxygen stoichiometry, with the suppression of superconductivity at $x>0.55$ and a metal-insulator crossover leading to a fully insulating phase when $x=1$ \cite{Maple_JSuper_1994, Radousky_JMatRes_1992, Akhavan_PhysB_2002, Booth_PRB_1994}.

    The effects of Pr substitution are multifaceted. The conduction electrons of Pr originate from $4f$ orbitals and in its trivalent form (the dominant form in Pr-YBCO \cite{Radousky_JMatRes_1992, Booth_PRB_1994}) the ionic radius of Pr is larger than that of Y. This directly disturbs the adjacent CuO$_2$ planes, essential for superconductivity.  Rather than altering the chain stoichiometry by a reduction of oxygen, the chains in Pr-YBCO are largely unaffected by Pr doping \cite{Neumeier_PhysicaC_1990}. Instead, hole doping in the planes is varied through two distinct mechanisms: orbital hybridization between the Pr$4f$ and O$2p$ orbitals in the CuO$_2$ plane leading to hole localization \cite{Fehrenbacher_PRL_1993}, and pair breaking due to the enhanced disorder in particular via spin exchange interaction between the holes and the magnetic moment of Pr ions \cite{Maple_JSuper_1994}. Recent ARPES measurements in Pr-YBCO challenge this point of view and suggest that the reduction of $T_c$ is due to the donation of electrons to the CuO$_2$ planes \cite{yang2025}.
    
    %, , and hole filling in the planes from a free Pr electron \cite{Fehrenbacher_PRL_1993,Maple_JSuper_1994}.

    % With doping $x$, all three lattice parameters grow monotonically while the ratio $b/a$, which is an indication of the level of the orthorhombic distortion, does not vary as significantly as YBCO \cite{Neumeier_PhysicaC_1990, Jorgensen_PRB_1990}.

    Considering these dissimilarities between the structure and doping of YBCO and Pr-YBCO, it is perhaps unexpected that the phase diagram of Pr-YBCO in Fig. \ref{fig:phasediagram} resembles closely that of pure YBCO \cite{Badoux_Nature_2016} comprising antiferromagnetic and superconducting domes with similar dependencies on their respective doping. Muon spin-relaxation and NMR have tracked the termination of AFM order from the Pr spins to $x = 0.4$ \cite{Cooke_PRB_1990, MacFarlane_PRB_2002} which lies within the superconducting dome. Similarly, AFM order within the SC dome has been detected by neutron spectroscopy in YBCO \cite{Haug_NJPhys_2010}.

%\cite{Tomkowicz_SSTech_1992, Radousky_JMatRes_1992, Maple_JSuper_1994}

    One crucial observation that has recently been reported in Pr-YBCO \cite{Kang_PNAS_2023,Martinelli2025} is the existence of short-range charge order which appears universal in the phase diagram of underdoped cuprates and intertwines strongly with SC and magnetic order \cite{Ghiringhelli_Science_2012,  Chang_NatPhys_2012, Wu_NatCom_2012, LeBoeuf_NatPhys_2013, Comin_Science_2014, Tabis_PRB_2017, Oliviero_npj_2024,Tranquada_PRL_2007, Fujita_PRB_2004, Frachet_NatPhys_2020}. In YBCO, charge order manifests as a plateau in the superconducting dome centred around a doping of $\frac{1}{8}$ \cite{Liang_PRB_2006,Koike_SSCommun_1991, Cohn_PRB_1999, Guguchia_PRL_2017}. Near this doping level there is a depression of the upper critical field $H_{c2}$ \cite{Grissonnanche_NatCommun_2014} and an enhancement of $T_c$ under the application of hydrostatic pressure \cite{Cyr-Choinière_PRB_2018} providing clear evidence for competition between SC and charge order. 
    %There is no clear evidence for an equivalent plateau in Pr-YBCO in $T_c$ \cite{Akhavan_PhysB_2002, Maple_JSuper_1994, Neumeier_PhysicaC_1990} or in $H_{c2}$ \cite{Jia_PRB_1992}, however an early zero-field transport study under hydrostatic pressure up to 20 kbar over the doping range $0\leq x\leq0.6$ found an enhancement in $T_c$ with a peak at $x=0.3$ at $P=6$ kbar \cite{Neumeier_PhysicaC_1988}, where charge order would be maximised in Pr-YBCO \cite{Kang_PNAS_2023}.
    In underdoped YBCO and Hg-based cuprates, the change of the dominant carrier type is demarcated by a sign reversal in the low-$T$ Hall effect \cite{LeBoeuf_PRB_2011, Badoux_Nature_2016,Doiron2013,Oliviero_npj_2024} and in the Seebeck coefficient \cite{Laliberte_NatComm_2011, Gourgout_PRX_2022}. The negative sign of both quantities observed at low doping levels is consistent with a Fermi surface containing at least a mobile electron pocket. The doping range over which this sign change appears \cite{LeBoeuf_PRB_2011,Badoux_Nature_2016} is roughly consistent with the regime in which NMR and x-ray measurements find a high-$T$ biaxial charge order in the CuO$_2$ planes, although it has been noted recently that the endpoint of CDW order as a function of hole is not easy to define \cite{Zhou2025}. Quantum oscillation measurements in underdoped YBCO  \cite{DoironLeyraud_Nature_2007, Yelland_PRL_2008, Bangura_PRL_2008} and Hg-based cuprates \cite{Oliviero_npj_2024,Barišić_NatPhys_2013,Chan_NatCommun_2016} have shown a low frequency oscillation alluding to a Fermi surface comprising small pockets. This is in stark contrast to overdoped Tl2201 which hosts a large hole-like Fermi surface \cite{Vignolle_Nature_2008}, as predicted by band structure calculations. Supporting theoretical investigations have found that the reconstruction of the large Fermi surface into small electron-like pockets is consistent with a Fermi surface reconstruction (FSR) by biaxial charge order \cite{Harrison_NPJ_2012, Oliviero_npj_2024}. 
    This biaxial charge order is short ranged with a correlation length up to 60 \AA ~and has half integer out of plane modulation. The wavevectors characterizing charge ordering in YBCO are incommensurate and slightly different along the $a$- and $b$-axes.
   
    Under the application of high magnetic fields \cite{Gerber_Science_2015,Chang_natcommun_2016}, uniaxial strain \cite{Kim_science_2018} and in thin film form \cite{Bluschke_natcommun_2018}, YBCO forms coherent 3D charge order which coexists with, and has the same incommensurate ordering vector and doping dependence as the 2D charge order. It has integer out of plane modulation. Measurements on detwinned crystals have shown that the 3D charge order modulations are uniaxial along the $b$-axis. The maximum achieved in-plane correlation length is about 200 \AA~at $B$ = 28 T~\cite{Jang2016}.
    
    In Pr-YBCO, signatures of 3D charge order were found using resonant x-ray scattering (REXS) at $x=0.3$ ($T_c = 50$ K) \cite{Ruiz_Natcom_2022}. These signatures were only observed at integer out of plane ($L$) modulation with correlation length exceeding 360~\AA~and no indication of the biaxial 2D charge order at half-integer values of $L$. 
    The intensity of the peak associated with the 3D charge order was detectable at room temperature and grew with temperature, reaching a maximum at $T_c$, similar to pure YBCO \cite{Vinograd_NatCommun_2024}. This was interpreted as a strong indication that 3D charge order in Pr-YBCO was stabilized by the extended dimensionality of the Pr$4f$ orbitals.
    A subsequent study \cite{Kang_PNAS_2023} using soft REXS on thin films of Pr-YBCO across the entire doping series found two charge order peaks at half-integer $L$ suggesting 2D charge order similar to pure YBCO, but not at integer $L$  e.g. without the stabilized 3D charge order. 
    More recently, a study combining grazing incidence x-ray diffraction and REXS measurements on thin films of Pr-YBCO at $x=0.3$ found two distinct charge orders with similar in-plane periodicity but with different value of $L$ \cite{Martinelli2025}. The peak at $L$=1 (attributed to 3D charge order in ref.~[\onlinecite{Ruiz_Natcom_2022}]) was interpreted as a Pr-related super-lattice structure, while the peak at $L$=0.5 was interpreted as evidence of biaxial 2D charge order.
    The 2D charge correlations were found at high temperatures with a short correlation length of $25-30$~\AA   -- 3 times smaller than that of pure YBCO \cite{Blanco-Canosa_PRB_2014}. Despite the discrepancies reported by different groups in Pr-YBCO, there is clear evidence of charge modulations in this system.

    % The charge correlations were found at high temperatures and the correlation length of the 2D charge order was $25-30$ \AA  - 3 times smaller than that of pure YBCO \cite{Blanco-Canosa_PRB_2014}.
    % The question is whether these short-range charge correlations suffice to reconstruct the FS in an inherently disordered system, such as Pr-YBCO.
    
    It is not obvious whether these short-range correlations would suffice to reconstruct the Fermi surface in Pr-YBCO.  To probe this, we have performed Hall effect and magnetotransport measurements on twinned Pr-YBCO crystals spanning the doping range $x=0.2-0.45$ and $T_c=42-65$~K. We find a $T$-dependent sign-change in $R_H$ which disappears with increasing Pr content (decreasing $T_c$) and is reminiscent of that in YBCO, indicating the presence of a density wave order and a concomitant Fermi surface reconstruction. We propose that this density wave order stems from charge correlations and that the apparent lack of competition with superconductivity is a consequence of the short correlation length \cite{Kang_PNAS_2023} in addition to high disorder levels. The doping level at which the reconstruction terminates coincides with the onset of the metal-insulator crossover from Pr spin at $x=0.4$ \cite{Cooke_PRB_1990}, highlighting the complex relationship between magnetic and charge order. Overall, we find stark similarities between the phase diagrams of Pr-YBCO and YBCO, and discuss the results in the context of disorder and the disparate mechanisms of doping holes into the CuO$_2$ planes.

\section{\label{Methods} Experimental Methods}

    Crystals of Pr-YBCO were synthesised in UC San Diego according to the method described in ref. [\onlinecite{Paulius1994}] and characterised using SQUID magnetometry. Gold contacts were sputtered onto polished crystals and annealed at 400$^\circ$C for two hours in flowing oxygen to allow the gold to diffuse. Electrical contacts were made using DuPont 4929N. The typical sample dimensions were on the order of $\sim200 - 500 \ \mu\mathrm{m}$ in length and breadth, and $30 \ \mu\mathrm{m}$ in thickness.  Pulsed magnetic field measurements up to 65 T were carried out at He-4 temperatures at the LNCMI-EMFL in Toulouse with two field polarities at each temperature to allow for a standard symmetrization and antisymmetrization of the magnetoresistance and Hall resistance, respectively. Note that $c$-axis contamination was observed in the resistivity $\rho_{xx}$ measurement but does not alter the Hall measurements due to the antisymmetrization. The six crystals, their nominal Pr content, and their $T_c$ from SQUID (onset) and dc resistivity (defined where $R = 0$) are shown in Table \ref{tab:samples}; the discrepancies between the two values are discussed below. To isolate the effect of Pr only, the chains were fully loaded with oxygen ($y=7$). Small deviations from optimal oxygen doping are possible, but are not responsible for the results reported herein. Due to the complex relation between Pr- and hole-doping and our inability to determine both the precise oxygen content and precise Pr content, it is not possible to label our samples with a specific hole doping $p$. We therefore use $T_c$ as a measure of sample doping and to facilitate comparison with YBCO.  

    \begin{table}[h!]
    \begin{tabular}{cccc}
    \hline \hline
        Sample & Nominal Pr $x$ & $T_c^{\mathrm{SQUID}}$ (K)  & $T_c^{\mathrm{res}}$ (K) \\ \hline 
        0.2  & 0.2  &  60  & 70 \\
        0.35a & 0.35 &  64  & 61 \\
        0.35b & 0.35 &  65  & 57 \\
        0.4  & 0.4  &  43  & 37 \\
        0.45a & 0.45 &  38  & 32 \\
        0.45b & 0.45 &  42  & 35 \\ \hline \hline
    \end{tabular}
    \caption{\label{tab:samples} Samples measured in this work with their nominal Pr content and transition temperature measured by SQUID and by resistivity. The discrepancy between the $T_c$ values is a possible consequence of inherent disorder and inhomogeneity in the system.}
    \end{table}

\section{\label{results} Results }
    \begin{figure}[t!]
        \includegraphics[width = 0.45\textwidth]{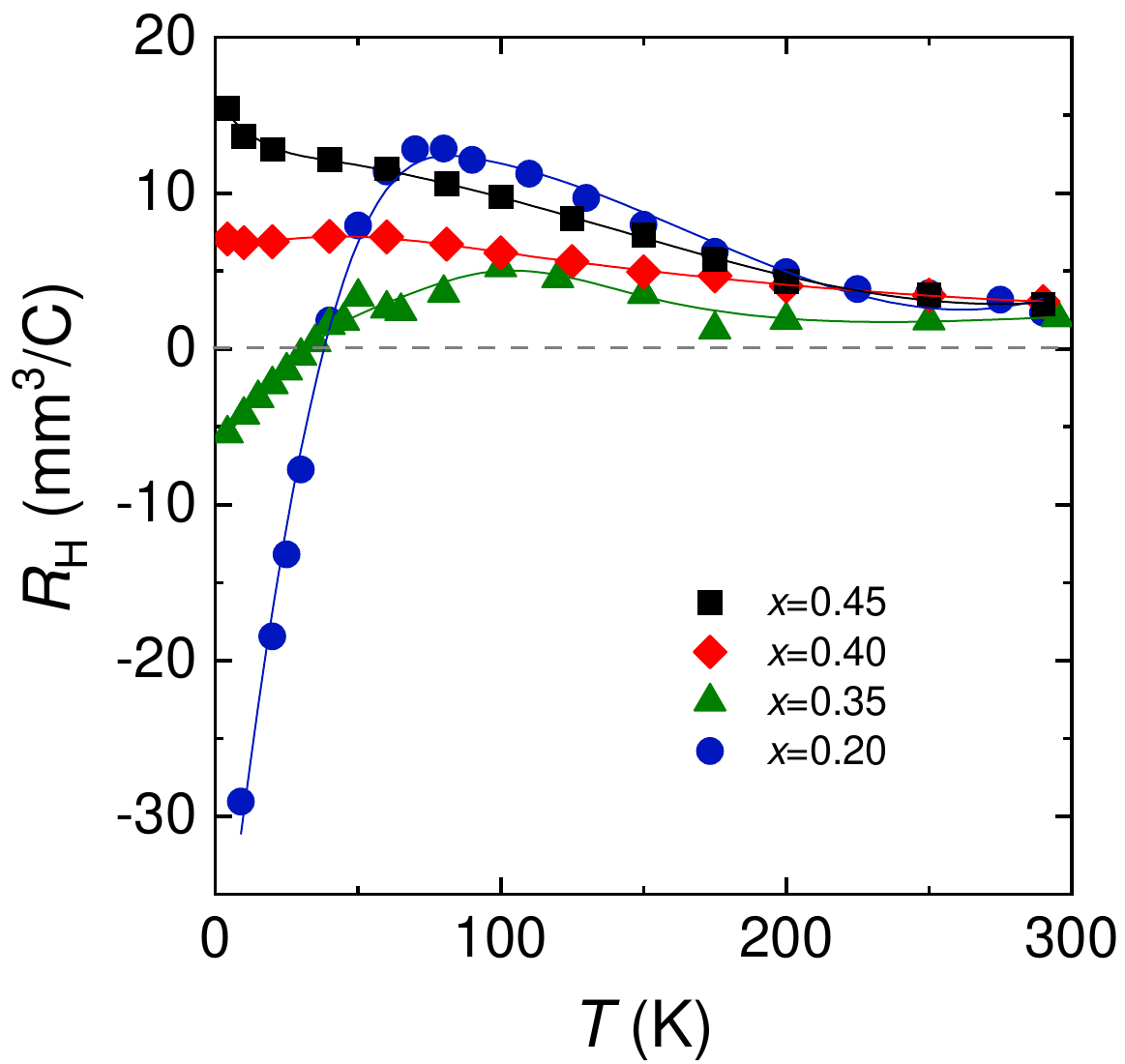}
        \caption{\label{fig:RH_max} $R_H(T)$ for four samples showing a change of sign at higher $T_c$. The temperature at which this occurs ($T_0$) decreases in parallel with $T_c$, eventually remaining positive for the low-$T_c$ crystals. The same behaviour is observed in the two samples which are not shown.}
    \end{figure}
    
    The Hall coefficient $R_H(T)$ at the maximum measured field strength (65~T) of four representative samples of Pr-YBCO is shown as a function of temperature in Fig. \ref{fig:RH_max}. The duplicates of $x=0.35$ and $x=0.45$ gave the same results as those shown and are not included in the figures for clarity. At high temperature where the Hall effect is perfectly linear in-field, we perform measurements up to 30~T and a linear extrapolation of $R_H$ up to 65~T is used. In all samples, $R_H(T)$ grows with decreasing $T$; below $T = 100 \ \mathrm{K}$, the different doping levels exhibit distinct behaviour. In the samples with the higher Pr content (lower $T_c$) $R_H$ grows with temperature and shows an upturn at low temperatures. In contrast, the samples with a lower Pr content (higher $T_c$) reach a maximum in $R_H(T)$ before sharply falling and changing sign, hence suggesting a FSR occurring in the doping phase diagram of Pr-YBCO outwith the region where the metal-insulator crossover occurs. $T_0$, the temperature at which $R_H=0$, decreases with increased Pr doping (decreasing $T_c$). This is our first main finding: the Hall coefficient of Pr-YBCO with varying $x$ mirrors the behaviour of pure YBCO with varying~$p$. 

\section{\label{comparison} Comparison with YBCO}    

\begin{figure}[h]
        \includegraphics[width = 0.45\textwidth]{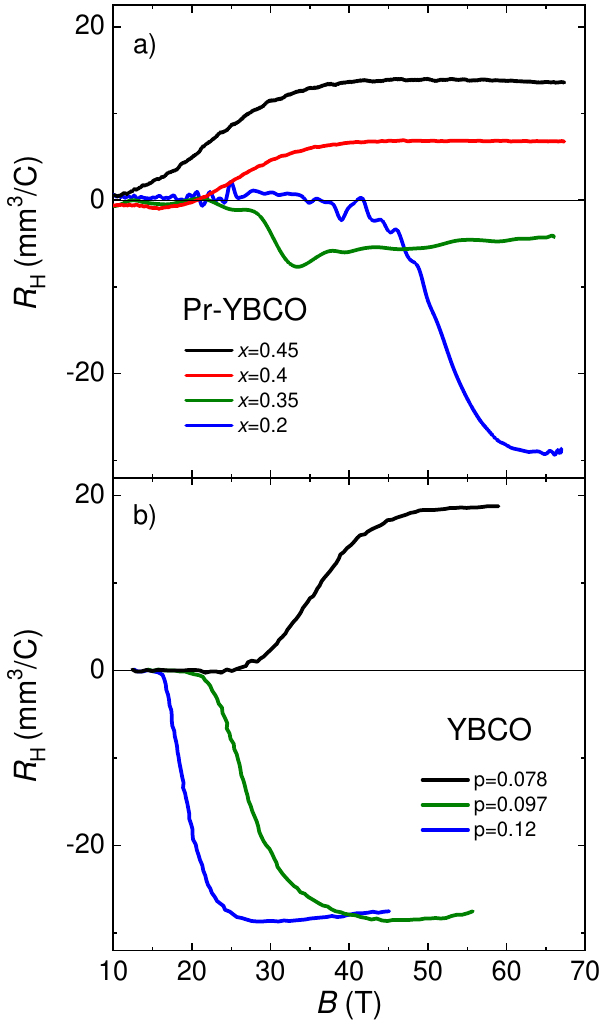}
        \caption{\label{fig:RH_10K} (a) $R_H$ as a function of $\mu_0 H$ at $T = 10$ K showing quite clearly the sign change that occurs and its doping dependence in Pr-YBCO. (b) The same plot but for pure YBCO at several doping levels (from ref. \cite{LeBoeuf_PRB_2011}). }
    \end{figure}

    To emphasize this point, in Fig. \ref{fig:RH_10K}, we plot $R_H$ versus magnetic field at $T$=10~K for (a) Pr-YBCO and (b) YBCO (from ref. \cite{LeBoeuf_PRB_2011}). The same general trends are evident in the two materials - at higher $T_c$, $R_H$ is negative whereas closer to the Mott state it is positive. As a reminder, in YBCO, the positive Hall effect at low doping ($p < 0.08$) has been interpreted as a consequence of a FSR by an antiferromagnetic  order. In this scenario, the FS consists of two hole pockets in the first Brillouin zone. This is in agreement with ARPES and QO studies showing the metallic character of the AFM phase at low doping in the IPs of a 5-layer cuprate \cite{Kunisada20}. At higher doping ($0.08 < p < 0.16$), the FS is reconstructed by charge order, presumably biaxial, which gives rise to an electron pocket and a negative Hall effect. In Pr-YBCO, the same phenomenology applies since for lower $T_c$, the Hall effect remains positive at $T=10$~K. This is justified by evidence for antiferromagnetic ordering of the Pr moments and the Cu moments within the Cu-O planes for $x > 0.4$ by µSr and NMR measurements \cite{Cooke_PRB_1990, MacFarlane_PRB_2002}. For the higher $T_c$ samples, the Hall effect changes sign at low temperature. 
    
\begin{figure}[h!]
        \includegraphics[width = 0.45\textwidth]{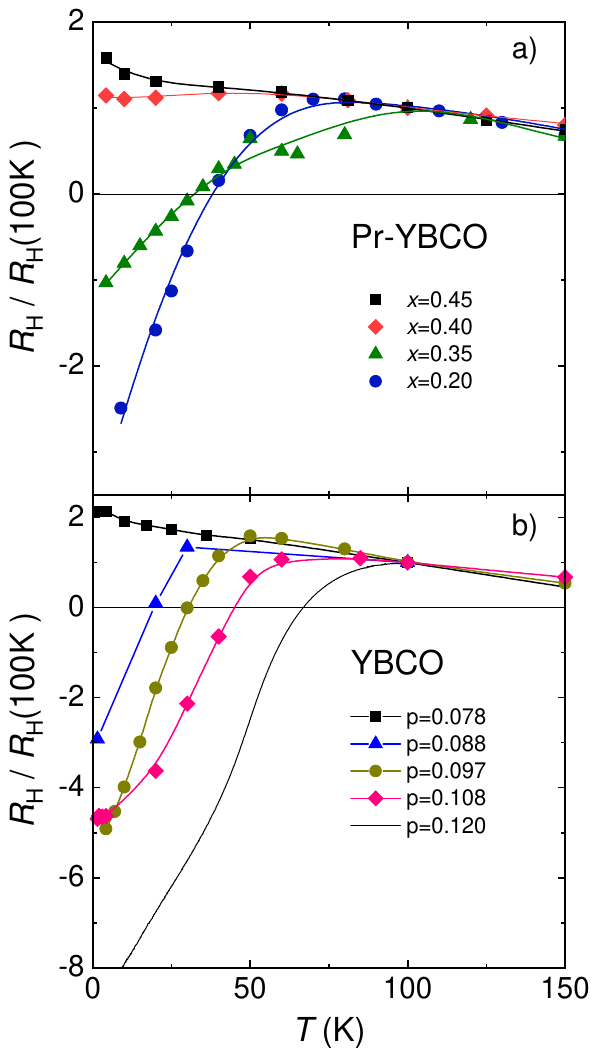}
        \caption{\label{fig:RH_T_dep} { $R_H(T)$ normalized by its value at $T=100$~K for (a) Pr-YBCO (this work) and (b) YBCO (from ref. \cite{LeBoeuf_PRB_2011}). Both show a systematic variation of $T_0$ with doping while the doping levels closest to the antiferromagnetic phase does not exhibit a sign change.}}
    \end{figure}

    In Fig.~\ref{fig:RH_T_dep}, we compare the temperature dependence of $R_H$ normalized by its value at $T=100$~K and measured at high fields for several dopings of (a) Pr-YBCO and (b) YBCO (from ref. \cite{LeBoeuf_PRB_2011}). Again, the similarity is striking. In YBCO, $R_H(T)$ goes from positive at $T=100$~K to negative as $T \rightarrow 0$, except for $p=0.078$, where $R_H(T)$ never changes sign and increases monotonically with decreasing temperature. In Pr-YBCO, $R_H(T)$ goes from positive at $T=100$~K to negative as T $\rightarrow$ 0 for $x=0.2$ and $x=0.35$, but never changes sign for $x=0.4$ and $x=0.45$.
    The sign reversal of the Hall effect is a signature of the presence of charge order and of an electron pocket in the reconstructed FS.
    
\begin{figure}[h!]
        \includegraphics[width = 0.45\textwidth]{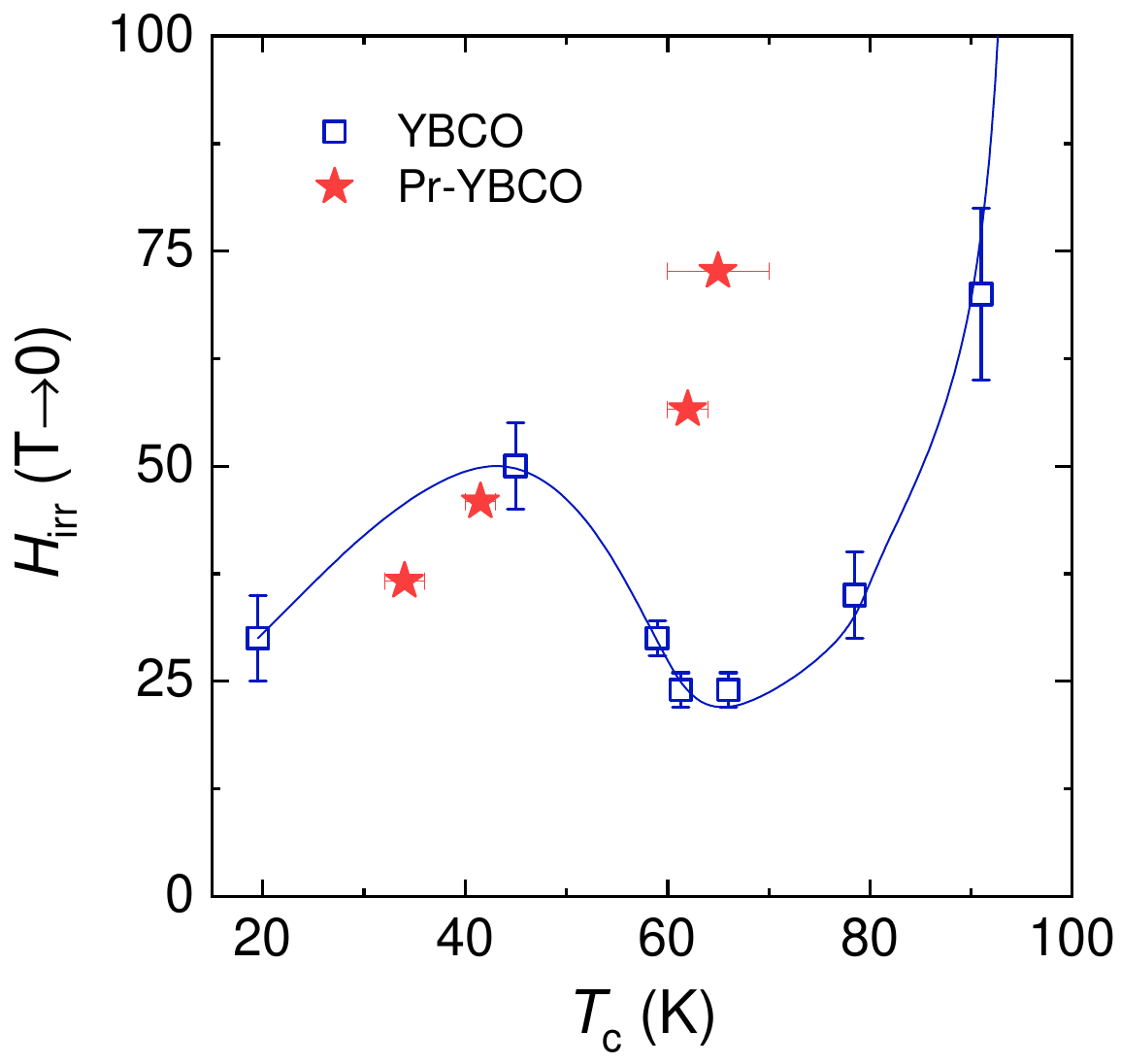}
        \caption{\label{fig:Hvs} {$H_{\rm irr}(T\rightarrow0)$ for YBCO \cite{Grissonnanche_NatCommun_2014} (blue squares) and Pr-YBCO (red stars) as a function of $T_c$. The clear depression seen in YBCO is not observed in Pr-YBCO representing the weaker competition between charge correlations and superconductivity in the latter. A monotonic dependence of $H_{c2}(0)$ in Pr-YBCO has also been observed in a more extensive doping range in ref.~\cite{Jia_PRB_1992}.}}
    \end{figure}

    One notable dissimilarity between the two materials is the doping-dependence of the irreversibility field $H_{irr}(T \rightarrow 0)$, which provides an estimate of the upper critical field, $H_{c2}$. In the Pr-YBCO system, an increase of Pr leads to a monotonic quenching of $H_{irr}$. The situation is more complicated in YBCO where $H_{irr}$ shows a clear suppression with increased hole doping, reaching a minimum near $p=0.125$ before increasing again towards $p^*$ \cite{LeBoeuf_PRB_2011,Grissonnanche_NatCommun_2014}.  The apparent lack of competition in Pr-YBCO in the zero-field resistivity \cite{Akhavan_PhysB_2002, Neumeier_PhysicaC_1990} and the magnetoresistance \cite{Jia_PRB_1992}, shown in Fig. \ref{fig:Hvs} poses a challenge to the reconstruction of the FS due to 2D charge order - if charge correlations do not compete measurably with superconductivity, are they able to cause a FSR? This hence leads to our second finding: the competition between superconductivity and 2D charge order in Pr-YBCO is far less pronounced than in YBCO.

\section{\label{discussion} Discussion}
\subsubsection{Doping mechanism}
    The substitution of Pr for Y is accompanied by subtle changes to the unit cell and a hybridisation between Pr$4f$ orbitals and O$2p$ orbitals, both of which affect the electronic character of Pr-YBCO. The larger ionic radius of Pr and its magnetic moment act as a source of disorder provoking a direct effect of Pr on the CuO$_2$ planes, hence on superconductivity and charge order.
    \newline
    Superconductivity in the cuprates is governed by the charge carriers in the CuO$_2$ planes. In pure YBCO the hole doping within the planes is governed by the oxygen stoichiometry of the CuO chains which act as charge reservoirs. 
    In Pr-YBCO, the dominant Pr ion (Pr$^{3+}$) localizes carriers in the CuO$_2$ planes as a consequence of orbital hybridization \cite{Fehrenbacher_PRL_1993}, but this interpretation has been challenged recently \cite{yang2025}. In the case of the former, the localized holes no longer contribute to the transport or superconducting properties and thus $T_c$ gradually falls with increased Pr content in the same way as $T_c$ falls as the CuO chains are diminished of oxygen. The depletion of superconductivity is further supplemented by pair-breaking effects due to disorder from the Pr. Note that none of the mechanisms which suppress superconductivity involve a change of the oxygen ordering in the CuO chains, and hence are entirely distinct from doping in YBCO. Nonetheless, in both YBCO and Pr-YBCO, the in-plane hole concentration is dictated by an out-of-plane influence and hence it is not unusual that the electronic behaviour associated with the planes be near analogous in the two systems. 

\subsubsection{Fermi surface reconstruction}
    On a similar vein, it is not unexpected that 2D charge order and a FSR are both present in Pr-YBCO, in particular, considering their universality in the cuprates \cite{Badoux_Nature_2016, Chan_NatCommun_2016, Oliviero_npj_2024,Laliberte_NatComm_2011}. Grazing incidence x-ray diffraction and REXS measurements on thin films of Pr-YBCO found charge order peaks at half-integer $L$ which have been interpreted as signatures of 2D charge order similar to YBCO \cite{Kang_PNAS_2023,Martinelli2025}. Furthermore, it is natural to conclude by comparison that the FSR reported here is a consequence of such 2D charge order. Although the correlation length is 3 times smaller than that of YBCO, we cannot exclude that it increases under a strong magnetic field, as it is the case in YBCO \cite{Gerber_Science_2015,Jang2016}.
    % The ordering wave vector measured in Pr-YBCO and YBCO are very similar. Therefore, it is natural to conclude that 2D biaxial charge order is present in Pr-YBCO, although the short correlation length poses a challenge to it being the cause of the FSR. However, we cannot exclude that the correlation length increases under a strong magnetic field, as it is the case in YBCO \cite{Gerber_Science_2015,Jang2016}.
   
    Unique to YBCO are uniaxial 3D charge correlations along the $b$-axis which appear under external influences such as stress \cite{Kim_science_2018,Vinograd_NatCommun_2024}, in thin films \cite{Bluschke_natcommun_2018}, and in magnetic fields exceeding 15 T \cite{Gerber_Science_2015,Chang_natcommun_2016,Laliberte_npj_2018}. The 3D charge order is superimposed on the 2D charge order, sharing the same incommensurate wave vector. The out-of-plane coherence is mediated by phase-locking of the CuO$_2$ bilayers. The in-plane correlation length can be as high as 200~\AA~at $B=30$~T, compared to 80-100~\AA~for the 2D charge order.
    \newline
    The exact mechanism of the FSR in YBCO by the CDW is still debated. Biaxial charge order with wavevectors ($Q_x$, 0) and (0, $Q_y$) creates a small electron-like pocket located at the nodes. This is the most natural scenario for the FSR, although it has been argued that the value of the coherence length of YBa$_2$Cu$_3$O$_{6.59}$ is too short to be at the origin of quantum oscillations at this doping level \cite{Gannot2019}. The 3D charge order has a longer coherence length but is uniaxial. A unidirectional charge order can create an electron-like pocket but only in the presence of nematic order and it would be located at the antinodes, where the pseudogap removes the density of states \cite{Yao2011}. There are other arguments against the scenario of FSR by 3D charge order. Since the incommensurate wave vector is identical to the one of the 2D charge order, it does not introduce a new periodicity. Although the onset of the 3D charge order is closer to the temperature at which $R_H$ changes sign, $T_0$, it was pointed out that for $p>0.12$, $T_0$ is larger than the onset of 3D charge order hence it is largely inert with respect to the FSR \cite{Laliberte_npj_2018}. This conclusion was also inferred from a transport study under uniaxial strain where the impact of 3D charge order was argued to be too weak to explain the sign reversal of the Hall coefficient as a function of temperature \cite{Nakata2022}. A recent REXS study on YBCO has shown that the stress induced 3D charge order is observed only in $y$ = 6.5 and $y$ = 6.67, while the sign reversal of the Hall effect occurs at least in the doping range between $y$ = 6.48 and $y$ = 6.86.
    
    It is not clear that Pr-YBCO hosts 3D charge order. The observation of signatures in REXS was supported by the argument that the 3D character of the hybridized Pr$4f$-O$2p$ orbital could facilitate long-range, out-of-plane charge order \cite{Ruiz_Natcom_2022}. However, a more recent study argued that the signature detected at $x$ = 0.3 is due to a super-lattice structure given by the 1/3 partial substitution \cite{Martinelli2025}. More work is needed to elucidate the nature of the 3D charge order in Pr-YBCO, in particular it would be interesting to perform x-ray scattering under magnetic field or uniaxial stress.
    
    % In Pr-YBCO, 3D charge order has been detected using REXS with large coherence length that is limited by the instrumental resolution. The proposition that Pr substitution into YBCO facilitates long-range, out-of-plane charge order relies on the 3D character of the hybridized Pr$4f$-O$2p$ orbital \cite{Ruiz_Natcom_2022}. In fact, it is not clear whether Pr-YBCO could host highly coherent 3D charge order akin to YBCO. It is already present at room temperature, but the sign reversal of the Hall effect occurs below $T_c$. Recently, this 3D charge order detected at $x$ = 0.3 was interpreted as a super-lattice structure given by the 1/3 partial substitution \cite{Martinelli2025}. More work is needed to elucidate the nature of the 3D charge order in Pr-YBCO, in particular it would be interesting to perform x-ray scattering under magnetic field.

\subsubsection{Effect of disorder}
    In archetypal CDW systems, the effect of disorder is to suppress and to smear the CDW transition. From a theoretical point of view, weak disorder prevents long-range incommensurate CDW order \cite{Nie2014}. But the signature of the ideal CDW may still appear in the presence of disorder. This has been demonstrated in Pd-intercalated ErTe$_3$, where two incommensurate CDW phases are progressively suppressed, as the amount of Pd increases \cite{Straquadine2019}.
    
\begin{figure}[t]
            \includegraphics[width = 0.45\textwidth]{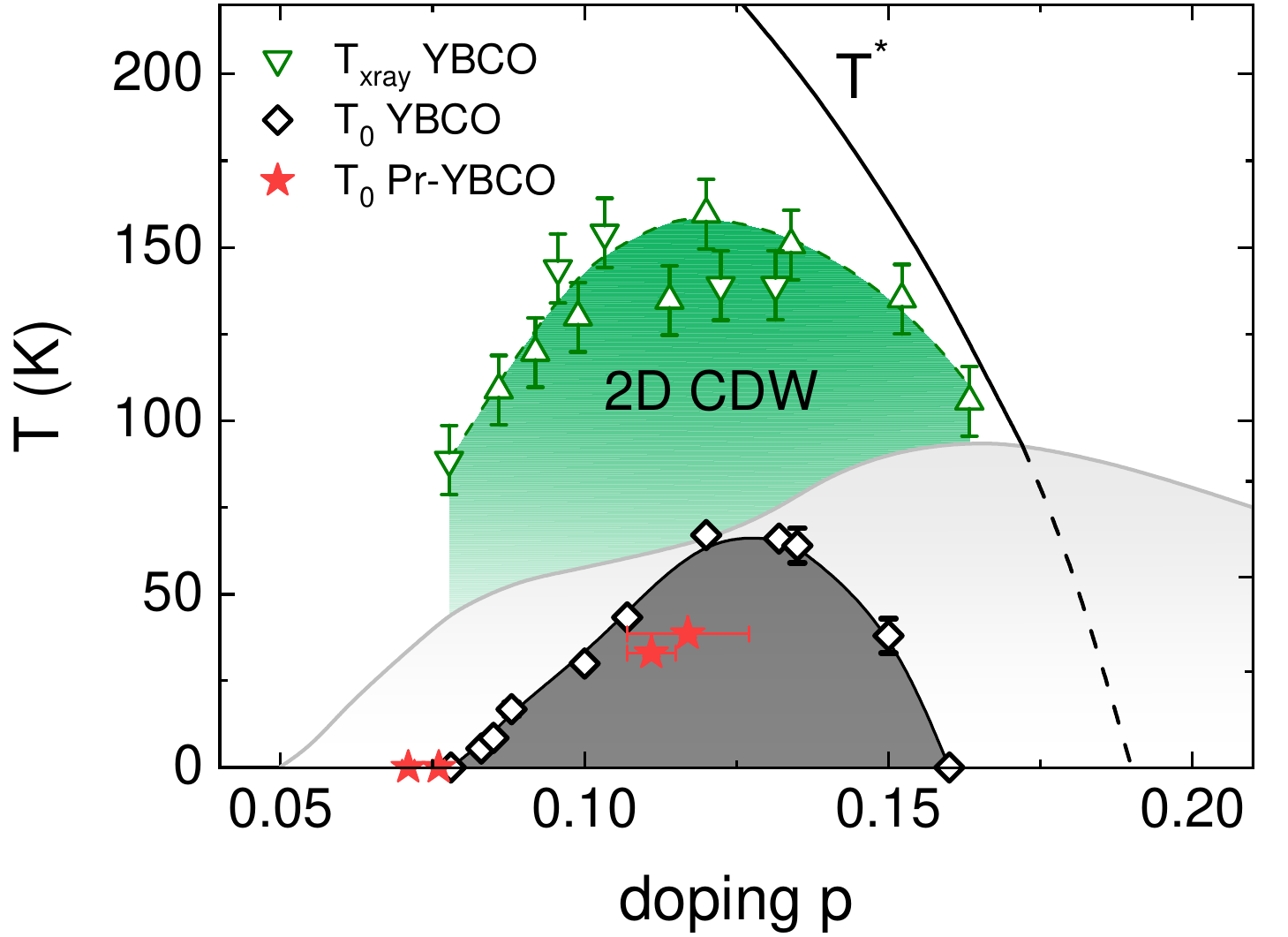}
            \caption{\label{fig:phasediagramybco} Comparison of the temperature of the sign reversal of the Hall effect, $T_0$ of YBCO (open diamonds) and of Pr-YBCO (red stars). The green region  marks the 2D charge order detected by x-ray measurements in YBCO (green triangles from refs.~\cite{Blanco-Canosa_PRB_2014,Hucker2014}). The superconducting dome is marked in gray. The solid black line and its extrapolation (dashed line) represent the pseudogap temperature in YBCO.}
    \end{figure}

    By analogy, we can understand why, for the same $T_c$, the sign reversal of the Hall effect occurs at lower temperature in Pr-YBCO compared to pure YBCO. This is shown in Fig.~\ref{fig:phasediagramybco} where the temperature of the sign reversal of the Hall effect of YBCO (open diamonds) is compared to Pr-YBCO (red stars).  Moreover, it is clear that the substitution of Pr results in a threefold decrease of the charge order correlation length \cite{Kang_PNAS_2023, Blanco-Canosa_PRB_2014, Bluschke_natcommun_2018}.
    This is also the case in Zn-doped YBCO, where Cu in the planes is substituted for Zn impurities, introducing point-like disorder. Similar to Pr-YBCO, $T_c$ is suppressed, with greater suppression at higher Zn percentages; but unlike Pr-YBCO, the decrease of $T_c$ is a direct consequence of pair-breaking and the doping is not modified by Zn substitution. Zn-YBCO hosts definitive charge correlations \cite{Blanco-Canosa_PRL_2013} and measurements (at $B=14$~T for Zn-YBCO) reveal a sign reversal of the Hall effect at a temperature $T_0$ which is a lower than pure YBCO, depending on the impurity level \cite{Abrami_Bristol_2021}. 
    Zn doping does not modify the charge order wavevector, but has a drastic effect on the correlation length and intensity; this was argued to be a consequence of suppressed charge correlations around the Zn impurities \cite{Blanco-Canosa_PRL_2013}. 
    
\subsubsection{Competition between CO and SC}
    
    A direct comparison with YBCO combined with the arguments presented above for Zn-YBCO suggests that 2D charge order in Pr-YBCO causes the FSR seen in Hall measurements. However, there is a lack of evidence in the data for the canonical competition between superconductivity and charge order, traditionally seen as a suppression of $T_c$ and $H_{c2}$ in transport measurements near $p\sim0.125$. It is pertinent to note that the reduced correlation length in Pr-YBCO alone cannot account for the lack of competition between charge order and superconductivity.
    % While the evidence for a FSR in Pr-YBCO is clear, the Hall effect results are used to intimate the existence of 2D charge order as its cause. However, there is a lack of evidence in the data presented above for the canonical competition between superconductivity and charge order, traditionally seen as a suppression of $T_c$ and $H_{c2}$ in transport measurements near $p\sim0.125$.    % Given the clear similarities between Zn- and Pr-YBCO and all the arguments discussed above, we can infer that the FSR in the latter is due to 2D charge order. It is pertinent to note that the reduced correlation length in Pr-YBCO alone cannot account for the lack of competition between charge order and superconductivity. 
    Indeed, in the clean single layer cuprate Hg1201, charge order has been reported with a similar correlation length as Pr-YBCO (7-8 lattice spacings) \cite{Tabis_PRB_2017} and yet a clear depression of $T_c$ remains manifest \cite{Yamamoto_PRB_2000}. Additionally, quantum oscillations find a single reconstructed pocket in Hg1201, signalling that the correlation length grows at high magnetic field and suffices to reconstruct the FS \cite{Barišić_NatPhys_2013,Chan_NatCommun_2016}. 
    
    If we assume that the correlation length grows at high magnetic field in order to reconstruct the FS, the concealed competition in Pr-YBCO is then also an effect of disorder. The disparate $T_c$ values of our samples raise the question of doping and/or Pr-substitution inhomogeneity.
    The suppression of the charge order and of SC in Pr-YBCO may not have comparable linear effects. Early pressure studies of the resistivity of Pr-YBCO up to 20 kbar do reveal an enhancement of $dT_c/dP$ in the doping region where we observe a sign change in $R_H$ \cite{Neumeier_PhysicaC_1988}. Similar results up to 2.5 GPa in YBCO have been interpreted as evidence for the enhancement of superconductivity once the charge order is suppressed by pressure \cite{Cyr-Choinière_PRB_2018}. The enhancement in Pr-YBCO is a factor of $\sim2$ smaller than in YBCO at the doping level where charge correlations are expected to be maximal. Additionally, above $P=6$ kbar, $dT_c/dP$ decreases at $x=0.3$ hinting that Pr hybridization has a complex effect on competing orders \cite{Neumeier_PhysicaC_1988}.

\section{\label{summary} Summary}

    Measurements of the Hall coefficient in Pr-YBCO at magnetic fields up to 65 T find that $R_H$ becomes negative in certain doping levels at low temperatures with the temperature at which the sign change occurs, $T_0$, diminishing with increased Pr doping (decreased $T_c$). $R_H$ remains positive at all temperatures at high $x$ where the normal state resistivity shows a metal-insulator crossover once superconductivity is suppressed. By analogy with other underdoped cuprates, we attribute the sign change of the Hall effect to a FSR from a large FS to small electron pockets due to 2D charge order. 

    Despite the differences in disorder and means of doping, the phase diagram of Pr-YBCO is qualitatively similar to pure YBCO. We speculate that this stems from the fact that the charge carriers within the planes are controlled by an out-of-plane influence in both systems: the CuO chain charge reservoirs in YBCO, and hole localisation in Pr-YBCO. The weakened competition of the 2D charge order with superconductivity, as evinced by the lack of plateau in $T_c$ and $H_{\mathrm{irr}}$, is a consequence of disorder and inhomogeneity, which effectively smear out well defined transitions. Our measurements cannot discern whether Pr-YBCO hosts 3D charge order similar to YBCO, yet this is an interesting question for future work to disentangle its origin, if present, and its connection to crystal structure, 2D charge order and superconductivity.
    
\section{\label{acknowledgments} Acknowledgments}
D.V. and C.P. acknowledges support from the EUR grant NanoX n\textsuperscript{o}ANR-17-EURE-0009 and from the ANR grant NEPTUN n\textsuperscript{o}ANR-19-CE30-0019-01. This work was supported by LNCMI-CNRS, members of the European Magnetic Field Laboratory (EMFL). Research at the University of California, San Diego, was supported by the US Department of Energy Basic Energy Sciences under Grant DE-FG02-04ER46105 (single crystal growth and characterization). A. F. acknowledges support from the National Science Foundation under Grant No. DMR-2145080.
\newline
\newline
$^{\dagger}$ caitlin.duffy@lncmi.cnrs.fr\\
$^\ddagger$ cyril.proust@lncmi.cnrs.fr\\
 
\bibliography{Pr_YBCO} 

\end{document}